\documentclass[aps,twocolumn,showpacs,floatfix,nofootinbib,10pt]{revtex4-1}
\usepackage{amssymb,amsmath}
\usepackage{graphicx,float,color}
\usepackage{indentfirst}
\newcommand{\be}{\begin{equation}}
\newcommand{\ee}{\end{equation}}

\newcommand{\ba}{\begin{eqnarray}}
\newcommand{\ea}{\end{eqnarray}}

\begin{document}
%
%
\title{The Minimal Geometric Deformation Approach Extended}
\author{R. Casadio}
\email{casadio@bo.infn.it}
\affiliation{Dipartimento di Fisica e Astronomia,
Alma Mater Universit\`a di Bologna,
40126 Bologna, Italy
\\
Istituto Nazionale di Fisica Nucleare, 
Sezione di Bologna, 40127 Bologna, Italy}
\author{J. Ovalle}
\email{Corresponding author: jovalle@usb.ve}
\affiliation{Abdus Salam International Center for Theoretical Physics (ICTP), Strada Costiera 11, Trieste 34014, Italy
\\
Dipartimento di Fisica e Astronomia, Alma Mater Universit\`a di Bologna, 40126 Bologna, Italy
\\
Departamento de F\'isica, Universidad Sim\'on Bol\'ivar, Apartado 89000, Caracas 1080A,
Venezuela}
\author{Rold\~ao da Rocha}
\email{roldao.rocha@ufabc.edu.br}
\affiliation{CMCC, 
Universidade Federal do ABC (UFABC) 09210-580, Santo Andr\'e, SP, Brazil.}
\begin{abstract} 
The minimal geometric deformation approach was introduced in order to study the exterior space-time
around spherically symmetric self-gravitating systems, like stars or similar astrophysical objects {as well},
in the Randall-Sundrum brane-world {{framework}}.
A consistent extension of this approach is developed here, which contains modifications of both 
the time component and the radial component of a spherically symmetric metric.
A modified Schwarzschild geometry is obtained as an example of its simplest application, and
a new solution potentially useful to describe stars in the brane-world is also presented.
\end{abstract}\maketitle
\flushbottom
%
%
%
\section{Introduction}
The idea that the universe we observe can be described by a sub-manifold (the brane)
embedded in a larger space (the bulk), has attracted great interest in physics.
From Kaluza and Klein's early attempts to the more recent models of extra-dimensions
by Arkani-Hamed, Dimopoulos and Dvali (ADD)~\cite{ADD,ADD1} and Randall and Sundrum
(RS)~\cite{RS}, what is ultimately sought after is a unified, hence simpler, description
of fundamental interactions.
While both the ADD and RS models could straightforwardly explain the hierarchy
of fundamental interactions, solely the RS model has a non-trivial bulk,
which makes it more attractive to explore gravity at high energies and
generalisations of four-dimensional General Relativity (GR).
Specifically, the new terms in the effective four-dimensional Einstein field equations
originating from the bulk, which can be viewed as corrections to GR, might be the key
to solve some open issues in gravity, like the dark matter problem~\cite{laszlo2011}. 
In this respect, from the phenomenological point of view, the search for solutions to
the four-dimensional effective Einstein field equations for self-gravitating systems is
particularly relevant, especially the case of vacuum solutions beyond the Schwarzschild
metric. 
\par
Unfortunately, even in the simplest case of the RS brane-world (BW), only a few
candidate space-times of spherically symmetric self-gravitating systems are known
exactly~\cite{dadhich,others,fabbri,wiseman,dejan,page,wiseman2}, mainly due to the complexity
of the effective Einstein field equations.
A useful guide in the search of such solutions is provided by the requirement that
GR be recovered at low energies.
Moreover, the RS model contains a free parameter, the brane tension $\sigma $,
which allows to restrain this prominent aspect by precisely setting the scale of high
energy physics~\cite{jovalle07}.
This fundamental requirement stands at the basis of the
Minimal Geometric Deformation (MGD) approach~\cite{jovalle2009},
which has made possible, among other things, to derive exact and physically
acceptable solutions for spherically symmetric~\cite{jovalle207} and non-uniform
stellar distributions~\cite{jovalleprd13} as well, to generate other physically acceptable
inner stellar solutions~\cite{jovalle07,jovalle2009,jovalle207,jovalleprd13,jovalleBWstars},
to express the tidal charge in the metric found in Ref.~\cite{dadhich}
in terms of the Arnowitt-Deser-Misner (ADM) mass, to study microscopic
black holes~\cite{covalle1,covalle2}, to clarify the role of exterior Weyl
stresses (arising from bulk gravitons) acting on compact stellar
distributions~\cite{ovallecqg2013}, as well as to extend the concept
of variable tension introduced in Refs.~\cite{GERGELY2008} by
analysing the behaviour of the black strings extending into the extra
dimension~\cite{cor2014} and to prove the existence of BW stellar
distributions made of regular matter with a vacuum Schwarzschild
exterior~\cite{ogc2015}.
In light of these results, it is interesting to study possible extensions of the MGD
approach.
\par
In addition, the field equations in the bulk and the brane were shown to be consistent
perturbatively~\cite{Casadioharms,bs1,rr2014,rr22014,Herrera-Aguilar:2015koa,kuerten2014}.
In this context, the four-dimensional solution is naturally embedded in the five-dimensional
bulk, with a black string-like object associated to the extended MGD procedure.
This result has already been established in the case of the standard MGD
technique~\cite{cor2014}. 
On a fluid brane with variable tension, the event horizon of the black string along
the extra dimension is a particular case of the bulk metric near the brane, being
based merely upon the brane metric. 
Numerical techniques shall be employed to study E\"otv\"os branes~\cite{Eotvos,Eotvos2}
in the extended MGD framework, showing the embedding of the solution in the five-dimensional bulk.
Such embedding represents a largely employed  method that has been previously
investigated~\cite{maartens,casadio1,Casadioharms} and moreover applied to a
variable tension brane~\cite{bs1}, incorporating inflation~\cite{bs1,rr2014,rr22014},
dark dust~\cite{Herrera-Aguilar:2015koa} and the realistic cases involving post-Newtonian
approximations for the Casadio-Fabbri-Mazzacurati black string~\cite{cfabbri,kuerten2014}. 
\par
In the next section, we shall review the standard MGD approach, which involves 
a modified radial component of the metric, and then extend it to the case
in which the time component of the metric is modified in section~\ref{III}.
In section~\ref{ext}, we shall describe the simplest application of the 
extended MGD to a Schwarzschild metric, and the relation between this MGD
extended solution on the brane and the associated five dimensional solution
in the bulk is discussed in section~\ref{secV}.
In section~\ref{secVI}, two new exact black hole metrics are presented as a direct
consequence of the extended MGD.
Finally, in section~\ref{secVII}, the possible extension of the MGD inside a self-gravitating
system is discussed in detail.
We summarise our work in section~\ref{conc}.  
\section{Minimal Geometric Deformation}
\label{II}
In the generalised RS BW scenario, gravity acts in five dimensions
and modifies the gravitational dynamics in the (3+1)-dimensional brane
accessible to all other physical fields.
The effective four-dimensional Einstein equations read~\footnote{We use
$k^{2}=8\,\pi \,G$, where $G$ is the 4-dimensional Newton constant, and $\Lambda$ is 
the 4-dimensional cosmological constant.} 
\begin{equation}
G_{\mu \nu }
=
-k^{2}\,T_{\mu \nu }^{\mathrm{eff}}-\Lambda \,g_{\mu \nu }
\ ,
\label{4Dein}
\end{equation}
and contain an effective energy-momentum tensor
\begin{equation}
T_{\mu \nu }^{\mathrm{eff}}
=
T_{\mu \nu }
+\frac{6}{\sigma }\,S_{\mu \nu }
+\frac{1}{8\pi }\,\mathcal{E}_{\mu \nu }
+\frac{4}{\sigma }\mathcal{F}_{\mu \nu }
\ ,
\label{tot}
\end{equation}
where $\sigma $ is the brane tension, 
\begin{equation}
S_{\mu \nu }
=
\frac{T\,T_{\mu \nu }}{12}
-\frac{T_{\mu \alpha }\,T_{\ \nu}^{\alpha }}{4}
+\frac{g_{\mu \nu }}{24}\left( 3\,T_{\alpha \beta}\,T^{\alpha \beta }
-T^{2}\right)
\end{equation}
represents a high-energy correction ($T=T_{\alpha }^{\ \alpha }$),
and 
\begin{equation}
k^{2}\,\mathcal{E}_{\mu \nu }
=
\frac{6}{\sigma }\left[
\mathcal{U}\left(u_{\mu }\,u_{\nu }+\frac{1}{3}\,h_{\mu \nu }\right)
+\mathcal{P}_{\mu \nu }+\mathcal{Q}_{(\mu }\,u_{\nu )}\right]
\vspace{0.0cm}\end{equation}
is a non-local source, arising from the five-dimensional Weyl curvature
($\mathcal{U}$ is the bulk Weyl scalar, $\mathcal{P}_{\mu \nu }$
and $\mathcal{Q}_{\mu }$ are the stress tensor and energy flux,
respectively), and $\mathcal{F}_{\mu \nu }$ contains contributions
from all non-standard model fields possibly living in the bulk. 
For simplicity, we shall assume $\mathcal{F}_{\mu \nu }=0$ and
$\Lambda =0$ throughout the paper.
\par
Let us now restrict to spherical symmetry, namely $\mathcal{P}_{\mu \nu}=\mathcal{P}\,h_{\mu \nu }$
and $\mathcal{Q}_{\mu }=0$,  and choose as the
source term in Eq.~(\ref{tot}) a perfect fluid, 
\begin{equation}
T_{\mu \nu }
=
(\rho +p)\,u_{\mu }\,u_{\nu }-p\,g_{\mu \nu }
\ ,
\label{perfect}
\end{equation}
where $\rho$ is the density, $p$ the pressure, and $u^{\mu }=e^{-\nu /2}\,\delta _{0}^{\mu }$
is the fluid four-velocity field in the Schwarzschild-like coordinates of the metric 
\begin{equation}
ds^{2}
=
e^{\nu}\,dt^{2}
-e^{\lambda}\,dr^{2}
-r^{2}\left( d\theta^{2}+\sin ^{2}\theta \,d\phi ^{2}\right)
\ .
\label{metric}
\end{equation}
Here $\nu =\nu (r)$ and $\lambda =\lambda (r)$ are functions of the areal
radius $r$ only, ranging from $r=0$ (the centre) to some $r=R$ (the surface)
inside the star, and from $r=R$ to some arbitrary $r$ in the outer vacuum,
where $\rho=p=0$.
\par
The metric~(\ref{metric}) must satisfy the field equations~\eqref{4Dein},
namely~\cite{covalle2} 
\begin{widetext}
\begin{eqnarray}
&&
k^{2}\left[ \rho
+\strut \displaystyle\frac{1}{\sigma }\left( \frac{\rho^{2}}{2}
+\frac{6}{k^{4}}\,\mathcal{U}\right) \right]
=
\strut \displaystyle\frac{1}{r^{2}}
-e^{-\lambda }\left( \frac{1}{r^{2}}-\frac{\lambda ^{\prime }}{r}\right)
\label{ec1}
\\
&&
\notag
\\
&&
k^{2}\strut \displaystyle\left[ p+\frac{1}{\sigma }\left( \frac{\rho ^{2}}{2}
+\rho \,p+\frac{2}{k^{4}}\,\mathcal{U}\right)
+\frac{4}{k^{4}}\frac{\mathcal{P}}{\sigma }\right]
=
-\frac{1}{r^{2}}+e^{-\lambda }\left( \frac{1}{r^{2}}+\frac{\nu ^{\prime }}{r}\right)
\label{ec2}
\\
&&
\notag
\\
&&
k^{2}\strut \displaystyle\left[ p+\frac{1}{\sigma }\left( \frac{\rho ^{2}}{2}
+\rho \,p+\frac{2}{k^{4}}\mathcal{U}\right) -\frac{2}{k^{4}}\frac{\mathcal{P}}{\sigma }\right]
=
\frac{1}{4}e^{-\lambda }\left[ 2\,\nu ^{\prime \prime}
+\nu ^{\prime 2}-\lambda ^{\prime }\,\nu ^{\prime }+2\,\frac{\nu ^{\prime}
-\lambda ^{\prime }}{r}\right]
\ ,
\label{ec3}
\end{eqnarray}
with primes denoting derivatives with respect to $r$.
Moreover, the conservation equation
\begin{equation}
p^{\prime }
=
-\strut \displaystyle\frac{\nu ^{\prime }}{2}(\rho +p)
\label{con1}
\end{equation}
holds unaffected.
The four-dimensional GR equations are recovered for $\sigma ^{-1}\rightarrow 0$,
and the conservation equation~(\ref{con1}) then becomes a linear
combination of Eqs.~(\ref{ec1})-(\ref{ec3}).
\par
From the above field equations, one finds that the radial metric component is 
in general given by
\begin{eqnarray}
e^{-\lambda}
&=&
\mu(r)
+\underbrace{e^{-I}\int^r\frac{e^I}{\frac{\nu'}{2}+\frac{2}{x}}
\left[H(p,\rho,\nu)+\frac{1}{\sigma}\left(\rho^2+3\,\rho \,p\right)\right]
dx+\beta\, e^{-I}}_{\rm geometric\ deformation}
\equiv
\mu(r)+f(r)
\ ,
\label{edlrwssg}
\end{eqnarray}
\end{widetext}
where $\beta=\beta(\sigma)$ is an integration constant, the function
\be
\label{I}
I(r,r_0)
\equiv
\int^r_{r_0}
\frac{\left(\nu''+\frac{{\nu'}^2}{2}+\frac{2\nu'}{x}+\frac{2}{x^2}\right)}
{\left(\frac{\nu'}{2}+\frac{2}{x}\right)}\,dx
\ ,
\ee
and $\mu=\mu(r)$ is the standard GR expression of the radial metric
component.
In particular, by assuming the space outside the star is empty, one
has
\begin{eqnarray}
\mu(r)
=
\begin{cases}
\label{intmass}
1-\strut\displaystyle\frac{2\,M}{r}
\ ,
&
\mbox{for}
\quad r>R
\\
\\
1-\strut\displaystyle\frac{k^2}{r}\!\!\int_0^r\! x^2\rho\, dx
\equiv
1-\frac{2\,m(r)}{r}
\,,
&
\mbox{for}\quad r\,\leq\,R
\ ,
\end{cases}
\end{eqnarray}
where $R$ is the radius of the star and $m=m(r)$ denotes
the standard GR interior mass function for $r<R$.
The constant $M$, in general, depends on the brane tension
$\sigma$ and must take the value of the GR mass $M_0 = m(R)$
in the absence of extra-dimensional effects.
\par
A crucial role in the original MGD approach is played by the function
$H$ in Eq.~\eqref{edlrwssg}, namely
\begin{eqnarray}
\label{H}
H
&\equiv&
-\left[ \mu ^{\prime }\left( \frac{\nu
^{\prime }}{2}+\frac{1}{r}\right)\right.
\left. +\mu \left( \nu ^{\prime \prime }
+\frac{\nu ^{\prime 2}}{2}+\frac{2\nu ^{\prime }}{r}+\frac{1}{r^{2}}\right)
-\frac{1}{r^{2}}\right]
\nonumber
\\
&&
+3\,k^2\,p
\ ,
\quad
\end{eqnarray}
which vanishes for any time metric function $\nu=\nu_{\rm GR}(r)$
that corresponds to a standard GR solution.
The geometric deformation in Eq.~\eqref{edlrwssg} is correspondingly ``minimal'',
that is simply given by contributions coming from the density and pressure
of the source.
One can thus start from a given GR solution $\nu=\nu_{\rm GR}(r)$,
then obtain the corresponding deformed BW radial metric function
$\lambda=\lambda(r)$ by evaluating the integral in Eq.~\eqref{edlrwssg}
with $H=0$, and finally compute the BW time metric function $\nu=\nu(r)$
from the remaining field equations.
It is worth noting that the correction $\nu-\nu_{\rm GR}$ necessarily
vanishes for $\sigma^{-1}\to 0$, and the starting GR solution is properly
recovered in this low energy limit.  
\par
Let us end this section by noting that the parameter $\beta=\beta(\sigma)$
in Eq.~\eqref{edlrwssg} could also depend on the mass $M$ of the self-gravitating
system and must be zero in the GR limit.
For interior solutions, the condition $\beta=0$ has to be imposed
in order to avoid singular solutions at the center $r=0$.
However, for vacuum solutions in the region $r>R$, where there is a Weyl
fluid filling the space-time around the spherically symmetric stellar
distribution, $\beta$ does not need to be zero, hence there can be a geometric
deformation associated to the Schwarzschild solution.
Overall, $\beta$ plays a crucial role in the search of BW exterior solutions,
and in assessing their physical relevance. In fact, we were previously able to constrain $\beta$ in
this MGD solution from the classical tests of GR in the solar system,
and the strongest constraint is $|\beta|\lesssim (2.80 \pm 3.45) \times 10^{-11}$,
from the perihelion precession or Mercury~\cite{submitted}.
\section{Extended Geometric Deformation}
\label{III}
BW effects on spherically symmetric stellar systems have already
been extensively studied (see,
e.g.~Refs.~\cite{kanti2013,matt2014,kuerten2014,debora2014,garcia2014,garcia1214,francisco2015,kanti2015} for some recent results and ~Refs.~\cite{wiseman,papa,creek} for earlier studies.
In particular, the exterior $r>R$ (where $\rho=p=0$) is filled with fields
(Weyl fluid) arising from the bulk, whose effects on stellar structures is not clearly
understood~\cite{ovallecqg2013}.
The MGD approach allows to study this region by generating modifications
to the GR Schwarzschild metric which, by construction, has the correct low energy limit. 
The next natural step is thus to investigate a generalisation of the
MGD for the exterior region filled with a Weyl fluid.
This can be accomplished by considering, in addition to the geometric deformation
on the radial metric component, given by the expression~\eqref{edlrwssg},
a geometric deformation on the time metric component, that is
\begin{equation}
\nu(r)
=
\nu_s+h(r)
\ ,
\label{def4}
\end{equation}
where $\nu_s$ is given by the Schwarzschild expression
\begin{equation}
e^{\nu_s}
=
1-\frac{2\,{M}}{r}
\ ,
\label{Schw00}
\end{equation}
and $h(r)$ is the time deformation produced by bulk gravitons,
which should be proportional to $\sigma^{-1}$ to assure the GR limit. 
Now, by using the expressions in Eq.~\eqref{edlrwssg}
(with $H=\rho=p=0$) and Eq.~\eqref{def4} in the vacuum equation
\begin{eqnarray}
R_\mu^{\ \mu}
&=&
e^{-\lambda}\left(\nu''+\frac{\nu'^2}{2}+2\,\frac{\nu'}{r}
+\frac{2}{r^2}\right)
\nonumber
\\
&&
-\lambda'\,e^{-\lambda}\left(\frac{\nu'}{2}+\frac{2}{r}\right)
-\frac{2}{r^2}
=0
\ ,
\label{vacuumgeneral2}
\end{eqnarray}
we obtain the following first order differential equation for the radial geometric
deformation $f=f(r)$ in Eq.~\eqref{edlrwssg}, in terms of the 
time geometric deformation $h=h(r)$,
\begin{equation}
 {\left(\frac{\nu'}{2}+\frac{2}{r}\right)}\,f'
+{\left(\nu''+\frac{{\nu'}^2}{2}+\frac{2\nu'}{r}+\frac{2}{r^2}\right)}\,f
+F(h)
=
0
\ ,
\label{defor}
\end{equation}
whose formal solution is given by
\begin{equation}
\label{genvacsol}
f(r)
=
e^{-I(r,R)}
\left(
\beta
-\int_R^r\frac{e^{I(x,R)}\,F(h)}{\frac{\nu'}{2}+\frac{2}{x}}
dx
\right)
\ .
\end{equation}
The exponent $I=I(r,R)$ is again given by Eq.~\eqref{I} and 
\begin{equation}
F(h)
=
\mu'\,\frac{h'}{2}+\mu\left(h''+\nu_s'\,h'+\frac{h'^2}{2}+2\,\frac{h'}{r}\right)
\ .
\label{F}
\end{equation}
The exterior deformed radial metric component is thus expressed as
\begin{eqnarray}
e^{-\lambda(r)}
&=&
1-\frac{2\,{M}}{r}
\nonumber
\\
&&
+
\underbrace{e^{-I(r,R)}
\left(\beta -\int_R^r\frac{e^{I(x,R)}\,F(h)}{\frac{\nu'}{2}+\frac{2}{x}}\,
dx\right)}_{\rm Geometric\ deformation}
\ .
\label{genvacsol2}
\end{eqnarray}
Finaly, by using the expressions in Eqs.~\eqref{def4} and~\eqref{genvacsol2}
in the field equations~\eqref{ec1} and~\eqref{ec2}, 
the Weyl functions ${\cal U}$ and ${\cal P}$ are written in terms of the radial
deformation $f(r)$ and time deformation $h(r)$ as
\begin{equation}
 \label{UU}
 \frac{6\,{\cal U}}{k^2\sigma}=-\frac{f}{r^2}-\frac{f'}{r}\ ,
\end{equation}
\begin{equation}
 \label{P}
 \frac{12\,{\cal P}}{k^2\sigma}=\frac{4\,f}{r^2}+\frac{1}{r}\left[f'+3f\frac{\mu'}{\mu}+3\mu\,h'+3fh'\right]\ .
\end{equation}
\par
To summarise, any given time deformation $h=h(r)$ will induce
a radial deformation $f=f(r)$ according to Eq.~\eqref{genvacsol2}.
In particular, a vanishing time deformation $h=0$ will produce $F=0$
and the corresponding geometric deformation will be again minimal.
For the Schwarzschild geometry, this procedure yields the deformed exterior
solution previously studied in Ref.~\cite{cor2014} (we note a constant $h$
also produces $F=0$, and corresponds to a time transformation 
$dT = e^{h/2}\,dt$).
\section{Modified Exterior Solution}
\label{ext}
A more interesting case is provided by non-constant time deformations
$h=h(r)$ such that $F(h)=0$, which will still produce a ``minimal''
deformation~(\ref{genvacsol}).
Let us therefore consider the non-linear differential equation
\begin{equation}
\label{Fzero}
F(h)=0\ ,
\end{equation}
whose solution is given by the simple expression
\begin{equation}
\label{Fsol}
e^{h/2}
=
a + \frac{b}{2\,M}\frac{1}{\sqrt{1-2\,M/r}}
\ ,
\end{equation}
where the integration constants $a$ and $b$ are both functions of the brane tension
$\sigma$.
Upon imposing the space-time is asymptotically flat, that is 
\begin{equation}
r\to\infty
\quad
\Rightarrow
\quad
e^{\nu}\to 1
\quad
\Rightarrow
\quad
h\to 0
\ ,
\end{equation}
one finds
\begin{equation}
a=
1 - \frac{b}{2\,M}
\ ,
\end{equation}
and the time metric component has the final form
\begin{equation}
\label{deftemp}
e^{\nu}
=
\left(1\!-\!\frac{2\,M}{r}\right)
\left[1\! +\! \frac{b(\sigma)}{2\,M}\left(\frac{1}{\sqrt{1-\frac{2\,M}{r}}}-1 \right)\right]^2
\ ,
\end{equation}
for $r>2M$, with the radial metric component given by
\begin{equation}
\label{minirad}
e^{-\lambda}=
1-\frac{2\,{M}}{r}
+\beta\, e^{-I}
\ ,
\end{equation}
where $I$ can be computed exactly but we omit it here for simplicity.
\par
A particularly simple case is given when $\beta=0$, 
which will produce no geometric deformation in the radial metric
component, so that $\lambda=-\nu_s$ is exactly the Schwarzschild 
form in Eq.~\eqref{Schw00}, and diverges for $r\to 2\,M$.
However, due to the modified time component~\eqref{deftemp}, 
this is now a real singularity, as can be seen by noting that 
the Kretschmann scalar $R_{\mu\nu\rho\sigma}R^{\mu\nu\rho\sigma}$
diverges at $r=2\,M$.
Moreover, this singularity is soft likewise,
since the higher order invariant
$(\nabla_\kappa\nabla_\tau R_{\mu\nu\rho\sigma})(\nabla^\kappa\nabla^\tau R^{\mu\nu\rho\sigma})$, 
 involving at least two derivatives of the curvature, further diverges at $r=2\,M$. 
The above solution could therefore only represent the exterior geometry of
a star with radius $R>2\,M$.
Since this solution  has no deformation in its radial metric component,
its dark radiation will be zero, ${\cal U} = 0$, as shown by Eq.~\eqref{UU}.
However, its Weyl function
\begin{equation}
\frac{{\cal P}(r)}{\sigma}
=
\frac{-b\,M^2\,k^2}{2\left[2\sqrt{1-\frac{2\,M}{r}}
-b\,M\left(\sqrt{1-\frac{2\,M}{r}}-1\right)\right]\,r^3}
\ ,
\end{equation}
which in fact diverges for $r\to 2\,M$.

Finally, Eq.~\eqref{deftemp} can be written for large $r$ as
\be
e^{\nu}
\simeq
1-\frac{2\,M-b}{r}-\frac{b\,(2\,M-b)}{4\,r^2}
\ ,
\ee
from which one can read off the ADM mass
\be
{\cal M}
=
M-\frac{b}{2}
\ee
and the tidal charge
\be
Q
=
\frac{b\,(2\,M-b)}{4}
\ ,
\ee
in terms of which the real singularity is located at
\be
r_c
=
2\,{\cal M}
\left(1-\frac{Q}{{\cal M}^2}\right)
\ . 
\ee
It would be now interesting to probe if one can change the nature of the
singularity at $r=2\,M$ if $\beta=\beta(\sigma,M)\not=0$.
In fact, the scalar $R_{\mu\nu\rho\sigma}R^{\mu\nu\rho\sigma}$ might
not diverge for $r\to 2\,M$ provided $\beta$ satisfies a very complicated algebraic
equation that depends on $M$ and $b=b(\sigma)$.
However, as we found that the classical solar system tests imply the very strong bound
$|\beta|\lesssim (2.80 \pm 3.45) \times 10^{-11}$, it is very unlikely
that $\beta$ meets this condition for arbitrary astrophysical masses $M$.
\section{Five-dimensional Solutions}
\label{secV}
Let $y$ denote the Gaussian coordinate, parameterising geodesics
from the brane into the bulk, where $n_Adx^A = dy$, for $A=0,1,2,3,5$ and $n_A$
the components of a vector field that is normal to the brane. 
The five-dimensional bulk metric ${g}_{AB}$ is related to the brane metric
$g_{\mu\nu}$ by ${g}_{AB}\,dx^A\, dx^B = g_{\mu\nu}(x^\alpha,y)\,dx^\mu\,dx^\nu + dy^2$.
In addition, the effective brane four-dimensional cosmological constant $\Lambda$,
the bulk cosmological constant $\Lambda_5$, and the brane tension $\sigma$ are fine-tuned
by~\cite{maartens}
$\Lambda=\frac{\kappa_5^{2}}{2}\left(\Lambda_{5}+\frac{1}{6}\kappa_5^{2}\sigma^{2}\right)$, 
where $\kappa_5=8\pi G_5$, where $G_5$ denotes the five-dimensional 
Newton gravitational constant. 
The brane extrinsic curvature, due to the junction conditions, reads 
\be\label{curv}
K_{\mu\nu}=-\frac{1}{2}\kappa_5^2 \left[T_{\mu\nu}+ \frac{1}{3}
\left(\sigma-T\right)g_{\mu\nu} \right]
\ .
\ee  
The bulk metric near the brane can be expressed as the Taylor expansion along
the extra dimension with respect to $y$ \cite{casadio1,maartens,bs1},
\be
g_{\mu\nu}(x^\alpha,y)
=
\sum_j
g_{\mu\nu}^{(j)}(x^\alpha,0)\,\frac{|y|^j}{j!}
\ ,
\ee
and one then finds that the expansion up to order $j=4$ given in Ref.~\cite{bs1}
is enough for a consistent analysis near the brane~\cite{bs1}.
For the sake of conciseness, we only display the metric up to the third order here,
that is
\begin{widetext}\ba
g_{\mu\nu}(x^\alpha,y)
&=&
g_{\mu\nu}(x^\alpha)
-\kappa_5^2\left[\frac{1}{3}(\sigma-T)g_{\mu\nu}+T_{\mu\nu}\right]\,|y| 
\nonumber
\\
&&
+\left[\frac{1}{2}\kappa_5^4\left(
T_{\mu\beta}T^\beta{}_\nu \!-\!\frac{2}{3} (T-\sigma)T_{\mu\nu}
\right)\!-\!2{\cal E}_{\mu\nu}\!-\!\frac{1}{3}\left(\!\Lambda_5\!-\!\frac{1}{6}
\kappa_5^4(T\!-\!\sigma)^2
\right)g_{\mu\nu}\right]\, \frac{y^2}{2!}
+\left.\Bigg[2K_{\mu\beta}K^{\beta}_{\;\,\rho}K^{\rho}_{\;\,\nu} - 
{\cal E}_{(\mu\vert\rho}K^{\rho}_{\;\,\vert\nu)}
\right.
\nonumber
\\
&&
\left.
\quad+ \frac{1}{6}
g_{\mu\nu}K\Lambda_5+R_{\mu\rho\nu\beta}K^{\rho\beta}-K{\cal E}_{\mu\nu}-\nabla^\rho{ B}_{\rho(\mu\nu)} 
-K^2K_{\mu\nu}+3K^\rho{}_{(\mu}{\cal
E}_{\nu)\rho}+K_{\mu\rho}K^{\rho\beta}K_{\nu\beta}
\Bigg]\;\frac{|y|^3}{3!}
\right.+\cdots
\nonumber
\label{tay}
\ea
\end{widetext}
where $g_{\mu\nu}(x^\alpha)= g_{\mu\nu}^{(0)}(x^\alpha,y=0)$, 
$R_{\sigma\rho\mu\nu}$ are the components of the brane Riemann tensor
-- here $R_{\mu\nu}$ and $R$ stand for the Ricci tensor and scalar curvature, respectively,
and ${B}_{\tau\rho\sigma}$ represents the trace free bulk Weyl tensor.
When the brane has variable  tension, the black string event horizon along the extra dimension
is affected and the above series contains additional terms~\cite{bs1}.
In particular, terms of order $|y|^3$ in the expansion~(\ref{tay}), involving derivatives of the
variable brane tension read~\cite{bs1}
\ba
g_{\mu\nu}^{(3)\,{}^{\rm extra}}
=
\frac{2}{3}\kappa_5^2\left[(\nabla_{(\nu}\nabla_{\mu)}\sigma
-
\Box\sigma)g_{\mu\nu}\right]
\ .
\label{addtruey3}
\ea
Furthermore, terms of order $y^4$ can be obtained straightforwardly,
although they are very cumbersome, and are given by Eq.~(12) of Ref.~\cite{bs1}.
In what follows, we shall only consider a time-dependent brane tension $\sigma = \sigma(t)$.
More details about spherically symmetric or anisotropic branes can be found in Ref.~\cite{alexca}. 
\par
Now, since the area of the five-dimensional horizon is determined by
$g_{\theta\theta}(x^\alpha,y)$~\cite{Casadioharms,maartens,bs1},
we consider the metric~(\ref{metric}) with the coefficients~(\ref{deftemp}) and~(\ref{minirad}).
In fact, $g_{\theta\theta}(x^\alpha,y=0) = R_H^2$ is precisely the black hole event horizon squared,
where $R_H$ denotes the coordinate singularity on the brane.
\par
In BW models, we can take into account the huge variation of temperature
in the Universe during the cosmological evolution.
In this context, fluid membranes of E\"otv\"os type~\cite{Eotvos} play a prominent role
in phenomenological aspects.
In fact, the so-called E\"otv\"os law states that the tension of the fluid membrane does
depend upon the temperature as $\sigma\sim \left(T_{critical}-T\right)$, 
where $T_{critical}$ denotes the highest temperature compatible with the existence of
the fluid membrane.
Hence, the BW is originated in an early Universe that is extraordinarily hot,  corresponding
to $\sigma \approx 0$.
The four-dimensional coupling constant $k$ and the brane tension $\sigma$ are considered
to be tiny in this context, which reinforces BW effects accordingly.
In a suitable thermodynamic framework~\cite{bs1,GERGELY2009}, we derived an
expression for the time-dependent variable brane tension $\sigma(t)=1-\frac{1}{a(t)}$,
where $a(t)$ denotes the scale factor in a Friedmann-Robertson-Walker universe,
given in our particular case by $a(t)\sim 1-\exp(\alpha t)$~\cite{bs1,cor2014}.
Moreover, in this context we further have ${\Lambda}\propto \left[1-{1}/{a(t)}\right]^2$~\cite{bs1},
which implies that $\Lambda$ has negative values that are led to small positive ones,
as the Universe expands.
Additional phenomenological analysis on this model was done in Ref.~\cite{bs1}. 
Hereupon, we consider $\Lambda_5 = 1=\kappa_5$ and the brane tension, with a lower bound 
$\sigma \sim 4.39 \times 10^8\,$MeV$^4$ to be normalized as well.
\par
\begin{figure}[h]
\begin{center}\includegraphics[width=3.2in]{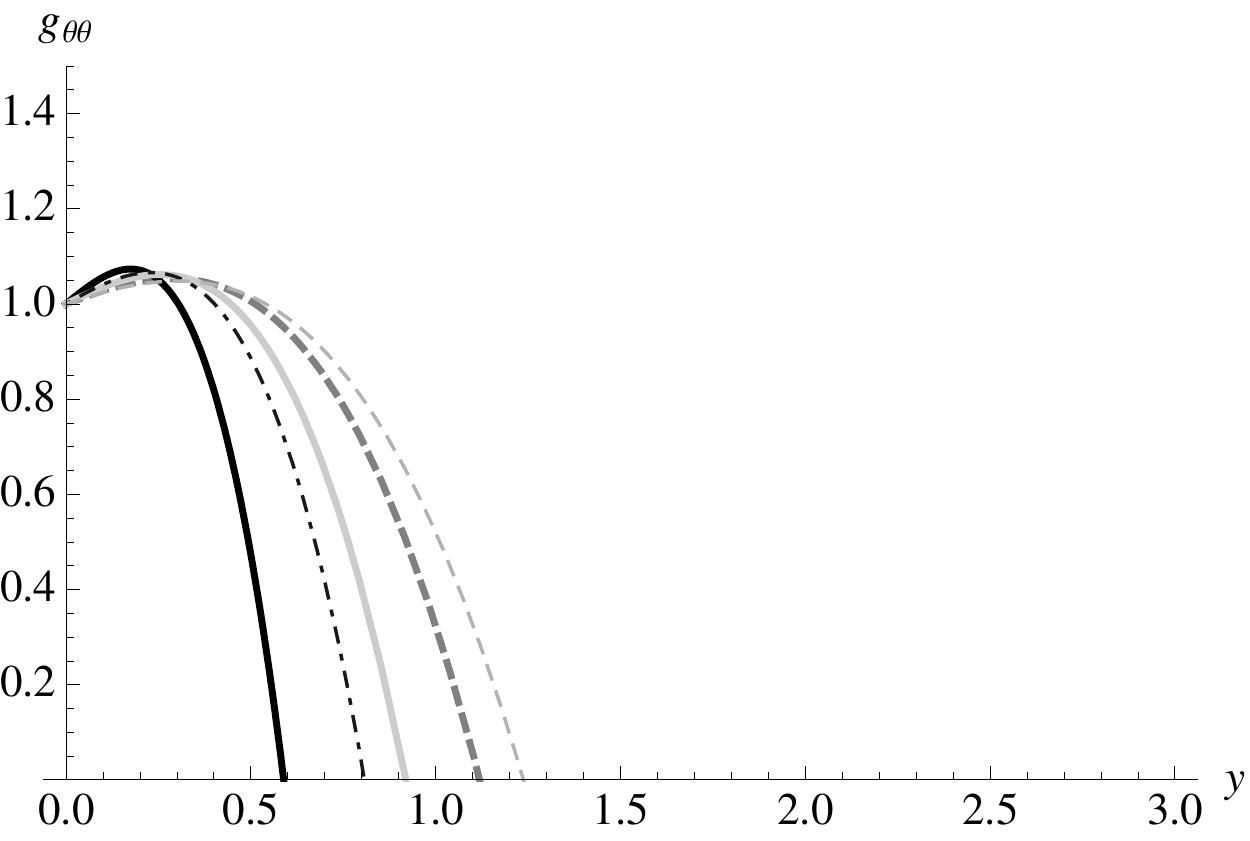}  
\caption{Squared black string horizon $g_{\theta\theta}(r=2\,M,y)$ 
 along the extra dimension
for $b(\sigma) = 1.9$ (solid black line), $b(\sigma) = 1.0$ (dash-dotted line), 
$b(\sigma) = 0.5$ (thick gray line), $b(\sigma) = 0.3$ (dashed gray line),
$b(\sigma) = 0.1$ (dashed light gray line).
Black hole mass $M=1$ and  $\beta\sim1/\sigma$.
 \label{fig2}}
\end{center}
\end{figure}

Below we plot the black string event horizon $r=2\,M$ for different values
of the brane tension $\sigma$.
In Fig.~\ref{fig2} the black string event horizon is shown to vary along the extra dimension as
a function of $\beta(\sigma)\sim \sigma^{-1}$ from the extended MGD technique.
It is worth mentioning that there is a point of coordinate $y_0$ along the extra dimension
where the horizon satisfies  $g_{\theta\theta}(r=2M,y_0)=0$. 
For instance, when $\sigma = 0.5$ then $y_0 = 0.88$, in Fig.~\ref{fig2}.
\par
\par
\begin{figure}[h]
\begin{center}
\includegraphics[width=3.2in]{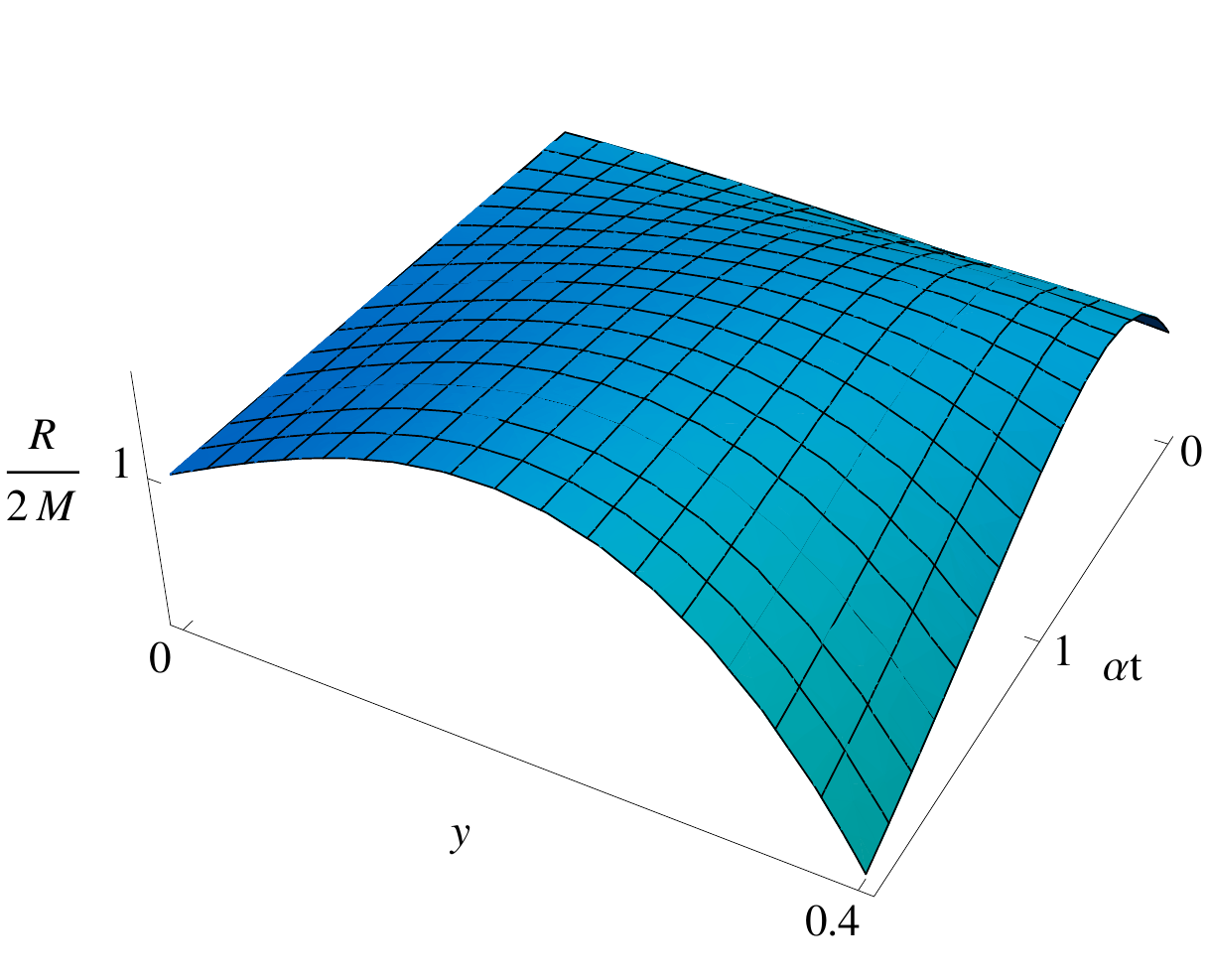}
\caption{Time-dependent black string event  horizon along the extra dimension,
for a variable brane tension. 
\label{fig1}}
\end{center}
\end{figure}
The plot in Fig.~\ref{fig1} exhibits the time-dependent black string horizon along the extra dimension.
As the cosmological time proceeds, the event horizon shrinks to a critical point along
the extra dimension, whose constant time slices have been already depicted in Fig.~\ref{fig2}
in a different range.  
These results are supported by the findings in Ref.~\cite{Horowitz:2001cz,kol},
and generalise those results to the extended MGD.  
From the perturbative analysis provided by the expansion~(\ref{tay}),
the black string horizon shrinks along the extra dimension.
On the other hand, black strings are linearly unstable to long wavelength perturbations.
If we perform the Gregory-Laflamme perturbation analysis,
we can show that, although in the range $0.4\lesssim y \lesssim 0.7$ the black string
horizon associated to the extended MGD shrinks, in other regions along the extra
dimension the event horizon's size can increase again, as for instance is shown in
Ref.~\cite{cor2014}.
Hence the total area is shown to increase, consistently with thermodynamic principles.  
A precursory numerical analysis reveals that the black string fragments along the extra dimension,
and forms five-dimensional black holes like those found in similar contexts~\cite{Lehner:2010pn}.
However, a complete numerical analysis adapted to the MGD extended framework goes beyond the
scope of the present paper and is thus left for future investigation.
\section{New Exact Solution}
\label{secVI} 
We will next investigate a more general solution for Eq.~\eqref{genvacsol2} by considering
a time deformation $h(r)$ producing $F(h)\neq\,0$.
A reasonable choice for $h(r)$ producing an analytical expression around the Schwarzschild
solution in the simple form 
\begin{equation}
\label{g00def}
 e^{\nu} = \left(1-\frac{2\,M}{r}\right)^{1+k}
\end{equation}
is provided by
\begin{equation}
\label{k-defor}
 h (r) = k\,\ln\left(1-\frac{2\,M}{r}\right)
 \ .
\end{equation}
We wish to emphasize that the parameter $M$ in Eq.~(\ref{g00def}) is not the ADM mass
${\cal M}$ of the self-gravitating system. 
Indeed they only coincide when there is not time deformation ($k = 0$). 
They are however related, as it is seen further below in equation~(\ref{ADM2}).
\par
Many exact configurations may be generated by using the ``deformation parameter''
$k$ in Eq.~(\ref{g00def}), the simplest one being the minimal geometric deformation 
associated to the Schwarzschild solution, which corresponds to $k=0$ (no time deformation). 
For $k=1$, Eqs.~\eqref{g00def} and~\eqref{genvacsol2} respectively yield 
\begin{equation}
 \label{k100}
 e^{\nu}\,=\,1-\frac{4\,M}{r}+\frac{4\,M^2}{r^2}
\end{equation}
and
\begin{equation}
 \label{k111}
 e^{-\lambda}\,=\,1-\frac{\left(2\,M-c_1\right)}{r}+\frac{\left(2\,M^2-c_1\,M\right)}{r^2}\ ,
\end{equation}
where
\begin{equation}
 \label{c1}
 c_1\equiv\,\frac{R}{1-M/R}\,\beta
\end{equation}
and $r_0 = R$, the radius of the distribution, was used to evaluate
the integral in Eq.~\eqref{I}.
\par
In order to obtain the asymptotic Schwarzschild behaviour
\begin{equation}
\label{Schw}
 e^{-\lambda}\,\sim\,1-\frac{{2\,{\cal M}}}{r} + {\cal O}(r^{-2})
\end{equation}
the constant $c_1$ must satisfy
\begin{equation}
 \label{c1f}
 c_1\,=\,-2\,M
 \ .
\end{equation}
Consequently, the expressions in Eqs.~\eqref{k100} and~\eqref{k111}
will reproduce the tidally charged solution given by~\cite{dadhich}
\begin{equation}
\label{casek1}
 e^{\nu}\,=\,e^{-\lambda}=1-\frac{2\,{\cal M}}{r}+\frac{Q}{r^2}\ ,
\end{equation}
where the ADM mass $ {\cal M} = 2\,M$ and the tidal charge $Q = 4\,M^2$.
The black hole solution in Eq.~(\ref{casek1}) corresponds to an extremal
black hole with degenerate horizons
\begin{equation}
r_h\,=\,{\cal M}
\end{equation}
which lies inside the Schwarzschild's radius.
Hence, according to this solution, extra-dimensional effects weaken
the gravitational field.
\subsection{New outer solution}
\begin{figure}[h]
\begin{center}\includegraphics[width=3.2in]{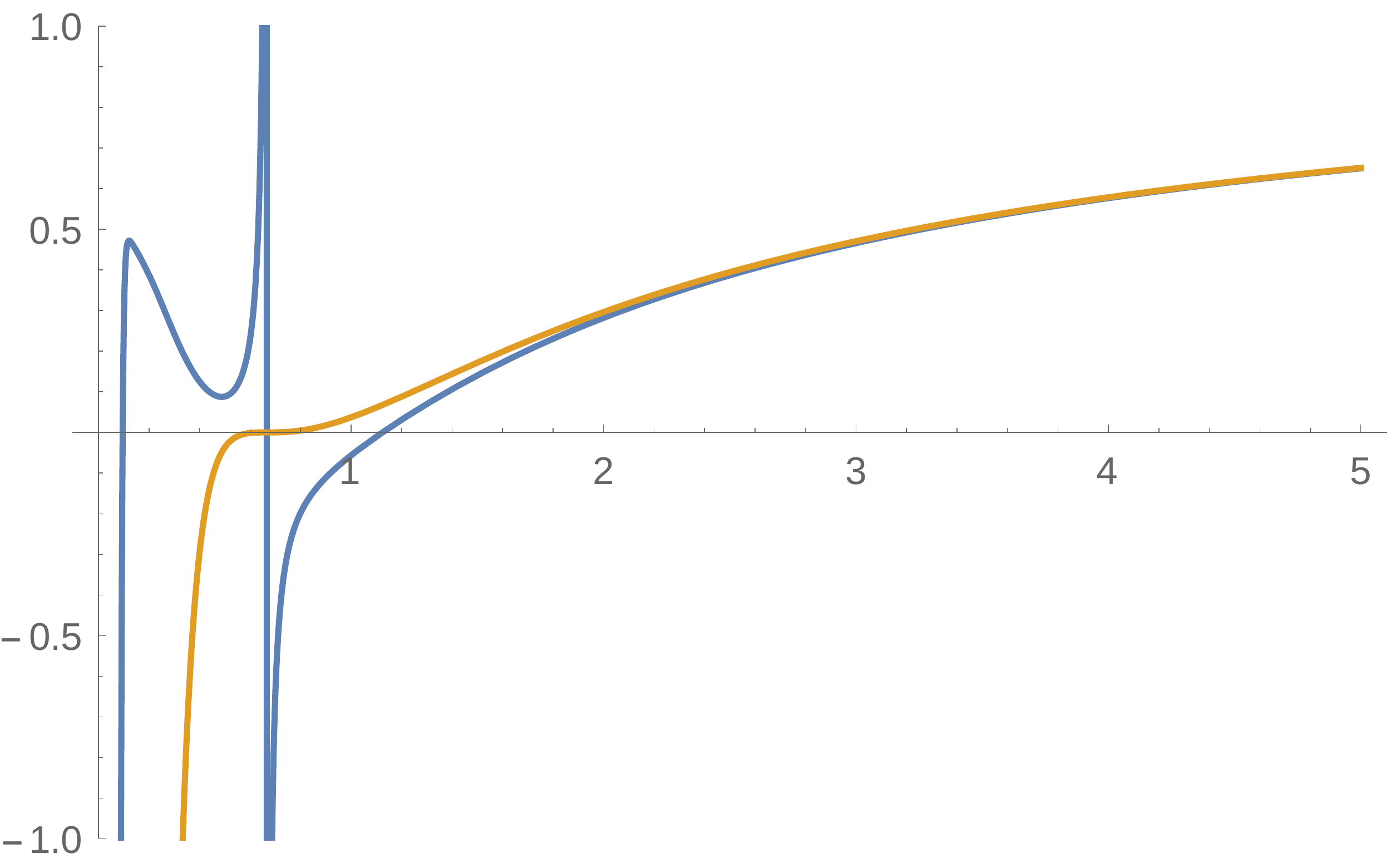}  
\caption{Behaviour of $g_{tt}(r)$ (red) and $g^{-1}_{rr}(r)$ (blue) for $k=2$.
We see two zeros and a singular point for $g^{-1}_{rr}$, the interior $r_i\simeq 0.095$, 
the middle $r_c=2/3$, and the exterior $r_e\simeq 1.124$.
It can be seen that the black hole horizon is shifted inside the Schwarzschild radius
($r_s=2\,{\cal M}$) by extra-dimensional effects.
The ADM mass is ${\cal M}=1$.
 \label{BH1}}
\end{center}
\end{figure}

Next we will show a new exact solution corresponding to $k=2$.
For $k=2$, Eqs.~\eqref{g00def} and~\eqref{genvacsol2} lead to
\begin{equation}
 \label{k200}
 e^{\nu}\,=\,1-\frac{2\,{\cal M}}{r}+\frac{Q}{r^2}-\frac{2}{9}\frac{{\cal M}\,Q}{r^3}\ ,
\end{equation}
and
\begin{widetext}
\begin{eqnarray}
 \label{k211}
 e^{-\lambda}
 =
 \left(1-\frac{2\,{\cal M}}{3\,r}\right)^{-1}&&\left[\frac{128\,c_2}{r}\left(1-\frac{{\cal M}}{6\,r}\right)^7+\frac{5}{224}\left(\frac{Q}{12r^2}\right)^4
 -\frac{5}{16}\frac{{\cal M}}{3r}\left(\frac{Q}{12r^2}\right)^3+\frac{5}{6}\left(\frac{Q}{12r^2}\right)^3\right.
\nonumber
\\
&&
\left. -\frac{25}{4}\frac{{\cal M}}{3r}\left(\frac{Q}{12r^2}\right)^2+\frac{25}{2}\left(\frac{Q}{12r^2}\right)^2
-\frac{5}{12}\,\frac{{\cal M}}{r}\frac{Q}{r^2}+\frac{10}{12}\frac{Q}{r^2}-\frac{4}{3}\frac{{\cal M}}{r}+1 \right]\ ,
\end{eqnarray}
\end{widetext}
where
\begin{equation}
\label{ADM2}
 {\cal M} = 3\,M;\,\,Q = 12\,M^2;\,\,\, c_2\equiv\,\frac{(1-2M/R)}{(2R-M)^7}\,R^8\,\beta\ .
\end{equation}
From the Schwarzschild limit in Eq.~\eqref{Schw}, we obtain 
\begin{equation}
c_2 = -\frac{{M}}{32}
\ .
\end{equation}
This solution displays two zeros of $g^{-1}_{rr}$, namely $r=r_i$ and $r_e$, and
a surface $r=r_c$ where $g^{-1}_{rr}$ diverges (and $g_{tt}=0$), all shown in Fig.~\ref{BH1}. 
These surfaces separate the space-time in four regions, namely
\begin{itemize}
 \item $0 < r < r_i$
 \item $r_i < r < r_c$
  \item $r_c < r < r_e$
 \item $r > r_e$.
 \end{itemize}
An exterior observer at $r > r_e>r_c$ will never see this singularity, as it is hidden behind 
the outer horizon $r=r_e$.
Nonetheless, we prefer to consider our solution as a candidate exterior of a self-gravitating
distribution with radius $R_o > r_c$, thus excluding the singular region $r = r_c$. 
Finally, since the exterior horizon lies inside the Schwarzschild radius, $r_e<r_s=2\,{\cal M}$,
this solution also indicates that extra-dimensional effects weaken the gravitational field.
\par
\begin{figure}[h]
\begin{center}\includegraphics[width=3.2in]{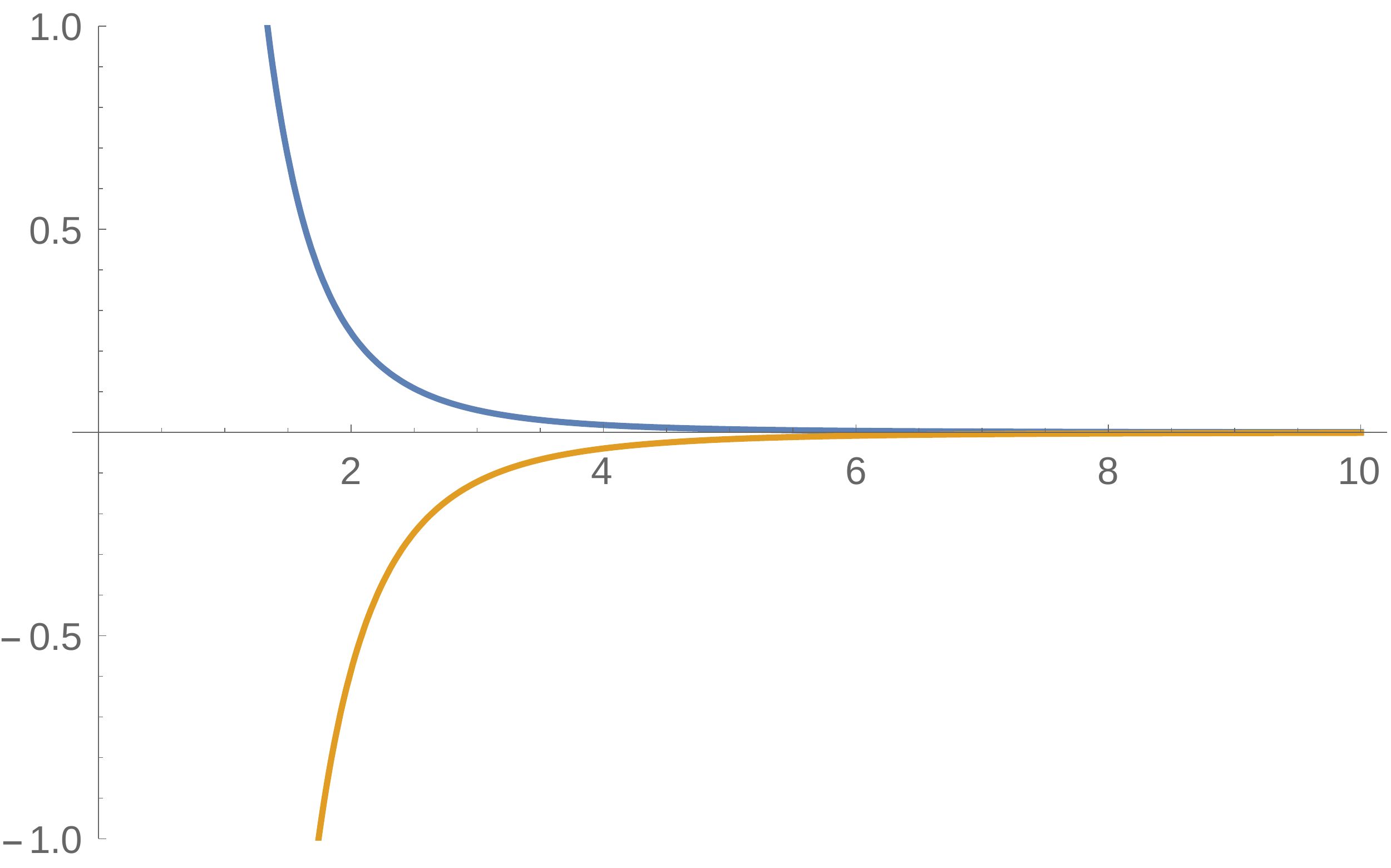}  
\caption{Behaviour of the Weyl function ${\cal U}(r)$ (upper curve) and ${\cal P}(r)$ (lower curve). The scalar Weyl function ${\cal U}$ 
increasing as we approach to the distribution until reach a maximun value inside the horizon, then it diverges as we approach to the critical value $h_c$. 
The opposite behaviour is seen for ${\cal P}$. The black hole mass is setting as ${\cal M}=1$.
 \label{UP}}
\end{center}
\end{figure}

Both Weyl functions ${\cal U}$ and ${\cal P}$ are shown in Fig.~\ref{UP}. 
We see that ${\cal U}$ is always positive, which indicates a negative 
radial deformation, according to Eq.~\eqref{UU}, and diverges at the 
singular surface $r=r_c$.
On the other hand, ${\cal P}$ is always negative, showing thus a negative
``pressure'' around the stellar distribution as consequences of extra-dimensional effects.
This function also diverges at the singular radius $r=r_c$.
\par
\par
\par
\subsection{The interior}
\label{secVII}
So far we have successfully developed an extension of the MGD
for the region $r>R_o$ surrounding a self-gravitating distribution.
The next logical step is to consider this extension inside the distribution $r<R_o$,
in order to investigate the consequences of bulk gravitons on physical variables
such as the density and pressure.
A critical point regarding the implementation of the MGD is the conservation
equation~\eqref{con1}, which contrary to GR, is not a linear combination of the
field equations~\eqref{ec1}-\eqref{ec3}.
We know that in the MGD approach any chosen GR solution will automatically
satisfy the conservation equation~\eqref{con1}.
However, when the time deformation in Eq.~\eqref{def4} is considered,
the conservation equation~\eqref{con1} becomes
\begin{equation}
p^{\prime }
=
-\strut \displaystyle\frac{\nu_0^{\prime }}{2}(\rho +p)-\strut \displaystyle\frac{h^{\prime }}{2}(\rho +p)
\label{consdeformed}
\end{equation}
where $\nu_0$ is the GR solution of the temporal metric component.
The expression in Eq.~\eqref{consdeformed} then yields 
\begin{equation}
0
=
-\strut \displaystyle\frac{h^{\prime }}{2}(\rho +p)
\label{consdeformed2}
\end{equation}
and therefore only a constant time deformation $h$, or equivalently a time transformation 
$dT = e^{h/2}\,dt$, may be implemented inside a self-gravitating system formed by a perfect fluid. 
The only way to overcome this problem is by considering a more complex interior structure,
like an anisotropic distribution, which leads to a more complex form of the conservation equation.
Furthermore, an exchange of energy between the bulk and the brane could be useful to 
implement the extension of the MGD inside an stellar distribution.
\section{Conclusions and outlook}
\label{conc}
The MGD deformation was consistently extended to the case when
both gravitational potentials, namely the radial and time metric
components, are affected by bulk gravitons.
We showed that the deformation for the time metric component
induces part of the deformation in the radial metric component, and
the latter can be minimised by assuming the former satisfies the
differential equation~\eqref{Fzero}.
\par
Using this extension of the MGD approach, a new possible exterior
geometry for a self-gravitating system was found.
This new solution presents a physical singularity at the Schwarzschild
radius $r=2\,M$.
Since there is no horizon, the singularity is naked, albeit not point-like,
and the geometry cannot be used to describe a BW black hole,
but might still be used to describe the exterior of a self-gravitating
star with size $R_o>2\,M$.
From the presently available observational data in the Solar system,
we were able to put constraints on one of the BW parameters. 
We also argued that the four-dimensional solution can be embedded in the bulk, 
as is supported by the black string analysis described in section~\ref{secV}.
We showed that the black string event horizon associated to the extended MGD
solution on the brane shrinks for the analysed values of the parameter that controls
the extension of the MGD procedure.
This is accomplished in the framework of a brane with variable tension. 
\par
Finally, a more general scenario was considered by allowing a time deformation $h(r)$
in Eq.~\eqref{def4} to affect the geometric deformation $f(r)$ in Eq.~\eqref{genvacsol},
in such a way that $F(h)\neq\,0$.
A corresponding deformation parameter $k$ was thus introduced which allows
to generate new exact exterior solutions.
In particular, the case $k=0$ represents no time deformation, as clearly seen
throughout the expression in Eq.~\eqref{k-defor};
$k=1$ represents the tidally charged metric corresponding to an extremal black hole
with degenerate horizons $r_h={\cal M}$;
the case $k=2$ represents a new exterior solution for a self-gravitating system
affected by the extra dimension in the form of a surrounding Weyl fluid.
This Weyl fluid has a positive effective energy density ${\cal U}$, which increases
as we approach the stellar distribution.
The effective pressure ${\cal P}$ is always negative, and decreases as we approach
the distribution, thus inducing anisotropy on the exterior near the self-gravitating
system.
Both functions tend to disappear rapidly as we move away from the stellar distribution,
and represent a ``Weyl atmosphere''.
\subsection*{Acknowledgments}
R.~Casadio is supported in part by the INFN grant FLAG.
R.~da~Rocha is supported in part by CNPq grants No. 303027/2012-6 and
No. 473326/2013-2, No. 451682/2015-7, and by FAPESP grant No. 2015/10270-0.
The research of J.~Ovalle is partially supported by the European Union
under Erasmus Mundus program, grant 2012-2646 / 001-001-EMA2,
and for the Abdus Salam International Center for Theoretical Physics, ICTP.
\end{document}